 \documentstyle{l-aa}
 \input epsf

\def\ïì{$\pm$}
\def\SLX1744{\mbox{SLX1744-299/300 }}
\def\*{$^{*}$}
\def\à{$^{a}$}

\def\1E1740{\mbox{1E1740.7-2942}}

\def\deg{$^\circ$}
\def\etal{{\it et~al.}}

\def\lesssim{\mathrel{\hbox{\rlap{\hbox{\lower4pt\hbox{$\sim$}}}\hbox{$<$}}}}
\def\gtrsim{\mathrel{\hbox{\rlap{\hbox{\lower4pt\hbox{$\sim$}}}\hbox{$>$}}}}

\begin{document}
\thesaurus{13(13.07.1)}

\title { Observations of the Soft Gamma-Ray early Afterglow emission from
Two Bright Gamma-ray Bursts.} 

\author{A.~Tkachenko\inst{1,2}, O.~Terekhov\inst{1,2},
R.~Sunyaev\inst{1,3},  R.~Burenin\inst{1}, 
C.~Barat\inst{4}, J.-P.~Dezalay\inst{4} and G.~Vedrenne\inst{4} } 

\institute{ 
Space Research Institute, Russian Academy of Sciences,
Profsoyuznaya 84/32, 117810 Moscow, Russia
\and
visiting Max-Planck-Institut f\"ur Astrophysik, Garching bei Munchen,
Germany 
\and
Max-Planck-Institut f\"ur Astrophysik, Garching bei Munchen, Germany 
\and
Centre d'Etude Spatiale des Rayonnements, Toulouse, France
}

\maketitle

\begin{abstract}

We present the  results of observations of the soft gamma-ray
early afterglows with the energy  above  $100~{keV}$ from two bright 
Gamma-ray bursts detected by the PHEBUS instrument of the  GRANAT
orbital  observatory.
We show that the light curves of
GRB~910402 and GRB~920723 events present the
afterglow emission with fading fluxes. During our
observations $(\sim700~{s})$ for these 
gamma-ray bursts the afterglow emission was fading as the power law 
of time
with indices equal to $-0.70\pm 0.04$ and $-0.60 \pm 0.05$ (at $1\sigma$
confidence level).
In both cases just after the end of the GRB event we observed 
the energy spectrum of the
afterglow  emissions which was
softer than the energy spectrum of the main GRB events.  
The average  photon index  of the main GRB event (in $100\--800~{keV}$ 
range) is equal to $\sim1.5$. In the beginning of  the afterglow emission
the  energy spectra of both events correspond to
the  much softer spectra with the photon index of $\sim2.5$.
We found that the times of abrupt softening of the burst spectra correspond
within statistical errors to the moment when the afterglow emission
 begins to dominate over GRB emission.
We found that during the afterglow emission of GRB910402 the  statistically 
significant  hardening of its spectra  was observed.
This is the first observation of hardening of GRB afterglow emission.
Analysis of GRB~910402 and its
afterglow   showed that this GRB source emits during $\sim 700~{s}$ 
 of our observations  in soft gamma-rays
$(100\--500~{keV})$  only $\sim1.6\%$ of 
its total energy released during the main event. For
the GRB~920723 we found that in afterglow during $\sim 700~{s}$ of
our observations  
$\sim6\%$ of the GRB total energy was released.


\end{abstract}

\markboth{A.Yu. Tkachenko et al.: Observations of Soft Gamma-Ray Afterglow}{}

\section{Introduction}
 
Recent  observations of the  X-ray, optical and radio afterglow emission
(e.g. Costa \etal\ 1997; Van Paradijs \etal\ 1997; Frail \etal\ 1997)
of Gamma-Ray Bursts as well as  measurements of the high  redshifts 
of the GRB sources in optical give a new impact to the gamma-ray burst
astronomy. It was shown that the sources of some of the detected GRB events
are located at cosmological distances.
X-ray and optical afterglow emission is fading as power law of time.
Such behaviour is consistent with the relativistic
fireball model of GRB (M\'esz\'aros\&Rees\ 1993, 1997).

The results of earlier observations indicated that afterglow might be
present right after gamma-ray burst events in x-rays (Sunyaev \etal\ 1990,
Murakami \etal\ 1991, Terekhov \etal\ 1993, Sazonov \etal\ 1998),  soft
gamma-rays (Klebesadel 1992, Tkachenko \etal\ 1995), gamma-rays
(Hurley \etal\ 1994).

The GRANAT observatory was launched into a high-apogee orbit
with the PROTON carrier rocketon December 1, 1989. Its four-day orbit with an
initial 200000~{km} apogee is of the kind that the satellite enters the Earth
radiation belts only for a  short interval (several hours). The satellite
was outside the Earth magnetosphere and the radiation
belts during 3 days in  every orbit. 
This ensures almost constant
background level during observations in absence of bright solar flares.
       
The PHEBUS instrument is the part of the payload of the observatory.
It consists of six cylindrical ($7.8~{cm}$ in diameter and $12~{cm}$ in
height)  $BGO$ detectors surrounded
by a plastic anticoincidence shield to reject the background
connected with the  charged particles.
The detectors were placed on
different sides of the GRANAT satellite, parallel to the axes of the
Cartesian coordinate system in such a way that with the probability
of $95\%$ 
at least two  detectors were able to observe a GRB event with no absorption
by the satellite mechanical structure. 
The instrument  was sensitive to 
photons in the  broad  $100~{keV}\--100~{MeV}$ energy range with the 
intrinsic total efficiency to gamma-rays equal to or greater $0.78$.
The field of view of the instrument was $4\pi~{ster}$.
Each of the six BGO detectors was equipped with a trigger system to detect
bursts. The trigger system activates electronics of the instrument to
transit to the ``burst mode'' automatically if the count rate
exceeds the background level by at least 8 standard deviations, in at
least two PHEBUS detectors. 

\begin{figure}[t] 
\centerline{\epsfxsize=76mm \epsffile{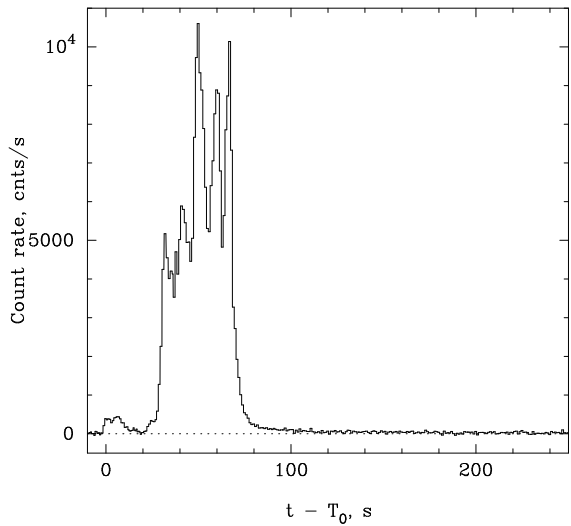}}
\vskip -5mm
\caption{\small The light curve of GRB~910402 in the $100{\--}500~{keV}$
energy range. $T_{0}=14^{h}27^{m}44^{s}.13~{UT}$.}
\vskip -5mm
\end{figure}

We present the results of observations of the soft gamma-ray
early afterglows with energy  more than $100~{keV}$ from two bright 
Gamma-ray bursts GRB~920723 and GRB~910402 detected by the PHEBUS
instrument. Both gamma-ray bursts  are strong events and give 
statistically significant count rate in  all 6 
detectors of the PHEBUS instrument.
GRB~910402 is the brightest burst observed by PHEBUS. The burst
Universal trigger time is $T_{0}=14^{h}27^{m}44^{s}.13~{UT}$.
Fig.1 shows the background-subtracted GRB~910402 light curve in
$100\--500~{keV}$ energy range. 

GRB~920723 has been detected by three instruments of the GRANAT
observatory: SIGMA, WATCH and PHEBUS in $8~{keV}\--24~{MeV}$
energy range (Terekhov \etal\ 1995).
For this burst the WATCH instrument  detected fading $8\--20~{keV}$
afterglow emission for more than 40~{s} after the end of the main
event (Terekhov \etal\, 1993).
Further analysis of the SIGMA data revealed soft  gamma-ray afterglow 
lasted for $\sim1000{s}$. 
The abrupt change of the GRB spectra was observed during transition 
to the afterglow emission in the $8\--200~{keV}$ energy range
(Burenin \etal\, 1999).
The PHEBUS burst Universal trigger time of this event is 
$T_{0}=20^{h}03^{m}08^{s}.3~{UT}$.
In Fig.2 the background subtracted  light curve of GRB920723 is
shown in $100\--500~{keV}$ energy range. 

BATSE did not see either of these events. GINGA has an event at
that time on 910402 and PVO did see GRB920723 (Fenimore 2000).

\begin{figure}[t] 
\centerline{\epsfxsize=76mm \epsffile{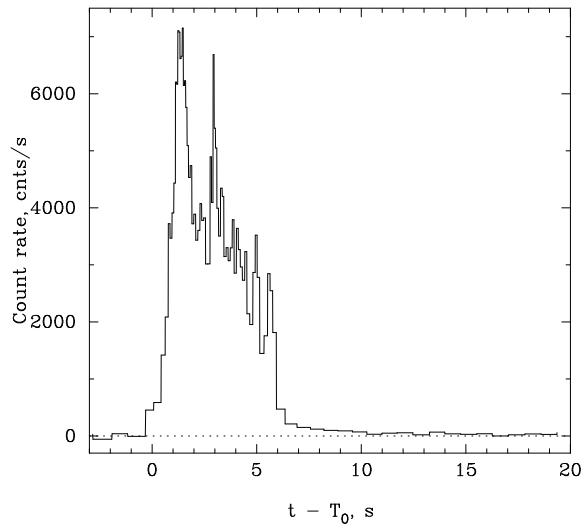}}
\vskip -5mm
\caption{\small The light curve of GRB~920723 in the $100{\--}500~{keV}$
energy range. $T_{0}=20^{h}03^{m}08^{s}.3~{UT}$.}
\end{figure}
\vskip -5mm

\section{Observation of the early afterglow emissions in GRB~910402 and
GRB~920723}

During detection of these two  gamma-ray bursts 
the GRANAT satellite was outside of the Earth
radiation belts and the Earth magnetosphere.
There were no solar flares during these gamma-ray bursts
(Solar Geophys. Data).
Detailed analysis of the gamma-ray light curves shows that the fluxes after
the  end of the main events  are not decreased to
the background level during several hundreds of seconds.
The background subtracted time histories of these two bursts
in $100\--500~{keV}$ energy range in
logarithmic scale are presented in the Fig.3 and 5.

Investigating behaviour of the time
history after
the end of the the main gamma-ray burst 
we found that in all 6 detectors the count rate 
is  varying around decreasing as the power law of time average values
with the standard deviation
correspondent to the Poissonian distribution.
The trend of the count rate after these two
bursts can be described by the same law~(\ref{first}) for
$t>T_1$.
 
\begin{equation}
\label{first}
N(t)=I (t-T_{1})^{\beta} 
\end{equation}

Using this simple  model it is possible to define 
intensity  of the afterglow emission  $I$ and
power law index of time  $\beta$ for each of six PHEBUS detectors.
It was also possible to estimate
$T_{1}$ - time when the afterglow emission fading as the power-law with time 
begins to dominate over the gamma-ray bursts emission. 

\begin{figure}[t] 
\centerline{\epsfxsize=75mm \epsffile{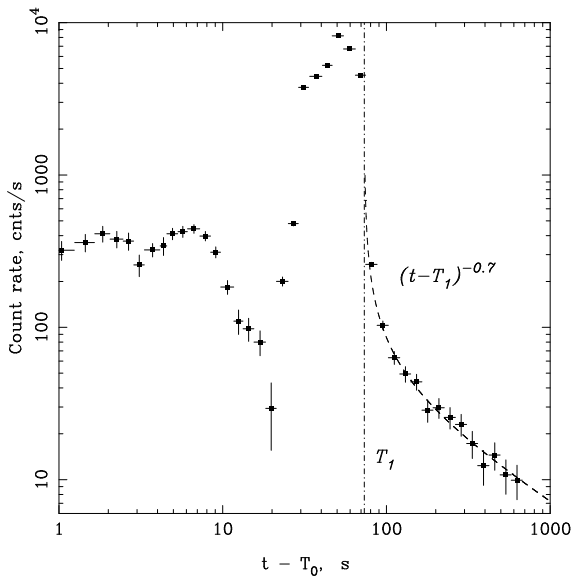}}
\vskip -5mm
\caption{\small The light curve of GRB910402 in $100\--500$~{keV} energy range in
 logarithmic coordinates. The instrumental background is subtracted. $T_{1}$ corresponds to the best fit estimate of the beginning of the afterglow emission. 
The dashed curve corresponds to the law $(1)$ for which the afterglow begins at
 moment $T_{1}$. Note that power line is curved in Log-Log space because the 
beginning of coordinates in this figure  ($T_{0}$) differs from moment 
$T_{1}$.}
\vskip -3mm
\end{figure}
\begin{figure}[th]
\centerline{\epsfxsize=76mm \epsffile{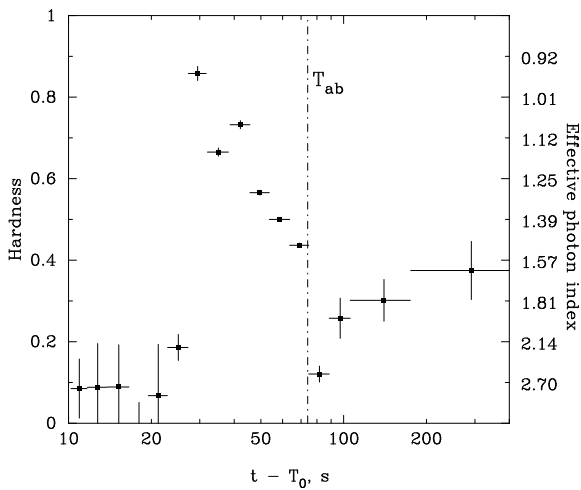}}
\vskip -5mm
\caption{\small The evolution of hardness of GRB~910402 (The ratio of the count
rates in $250\--800~{keV}$ to $100\--250~{keV}$ energy ranges). The values of
effective photon index on the right axis 
corresponds to the hardness values (left axis) for the power law spectra.
$T_{ab}$ corresponds to the moment of the  abrupt softening of the emission
from GRB source.}
\vskip -5mm
\end{figure}

We have found that the  excessive (over background level in each of the
six PHEBUS BGO detectors) count 
rates $I$ were proportional to the intensities
of the main GRB events in each detector as it had been expected in the case 
if the afterglow emission
is really connected with gamma-ray burst source.
For each of the bursts the  $\beta$  parameter was equal within 
statistical errors  for all
detectors of the instrument. 

We have found that for
GRB910402 the best fit corresponds to 
$\beta=-0.70 \pm 0.04$ (at $1\sigma$ confidence level)
and $T_{1}=73.2~{s}$ (${\chi}^{2}/dof=2.4/10$). The $1\sigma$ confidence 
interval
for $T_{1}$ is from 62.2 to 77.9~${s}$.  
To obtain these parameters we have used the afterglow count rate data
from 85 to 700~${s}$ after $T_{0}$.

For GRB920723 the best fit 
corresponds to 
$\beta=-0.60 \pm 0.05$ (at $1\sigma$ confidence level)
and $T_{1}=6.1~{s}$ ($\chi^{2}/dof=4.8/9$). The $1\sigma$ confidence interval
for $T_{1}$ is from 3.5 to 7.3~${s}$  
To obtain these parameters we have used the afterglow count rate data from 
8 to 700~${s}$ after $T_{0}$.
 
The energy spectra of the main bursts and afterglow emission are
different.
In both cases (GRB~910402 and GRB~920723) just after GRB event the energy 
spectrum of the
afterglow  emissions is much
softer than the energy spectrum of the main GRB events. 
Fig. 4 and 6 show the evolution of the spectral hardness of
the bursts.
We define spectral hardness as the ratio of count rate in $250\--800~{keV}$
to $100\--250~{keV}$ count rate)
We define the effective photon index 
as the index of a power-law spectrum
(in the $100\--800~{keV}$ energy range) which, would produce
the observed hardness at an incidence angle of 90\deg to the detector
axis. 

At the end of the both bursts the spectral hardness
ratio is equal to $\sim0.4\--0.5$ (fig.4 and 6). This corresponds to
the photon index of $\sim1.3\--1.5$  at the same energy range.
For the GRB~910402 during the time interval less than
$\sim10~{s}$ we observe drop of the hardness down to the value
of $\sim 0.1$ (fig.4). This value corresponds to the
photon index of $\sim2.5\--2.6$.
The moment of abrupt drop of the hardness
corresponds to $T_{ab}=74\pm10~{s}$ for GRB910402 (fig.4).
For the burst
GRB920723 the time of the abrupt drop of the spectral
hardness down to $\sim0.1\--0.2$ corresponds to
the moment $T_{ab}=6\pm1~{s}$ (fig.6). 
Thus the times of abrupt softening of the burst spectra
$T_{ab}$ coincide within statistical errors with the  ($T_{1}$) -  beginning
of the power law emission   fading during afterglow  for both GRB~910402
and GRB~920723. 
 
Note that as one can see from fig.4 during the afterglow emission of 
the GRB910402 a statistically significant hardening of the spectra
is observed. In the case of GRB~920723 the error bars are
too large (fig.6) to make any conclusions about behaviour of the
afterglow spectra.

The afterglow emission intensities in $100\--500~{keV}$ energy range 
are rather faint in comparison with 
gamma-ray bursts.  
Just a few tens of seconds after 
gamma-ray bursts discussed in this paper the afterglow intensities 
are $\sim1\%$ of the GRB maximum intensities.
In the case of GRB~920723 the  afterglow emission contains 
$\sim 6\%$ of the total burst energy
that was  emitted   during the interval of $6.1\--700~{s}$
after $T_{0}$ in the $100\--500~{keV}$ energy range. 
In the case of GRB~910402 only  $\sim1.6\%$  of the GRB energy 
was released during the observed afterglow  ($73.2\--700~{s}$
after $T_{0}$).

\begin{figure}[t] 
\centerline{\epsfxsize=75mm \epsffile{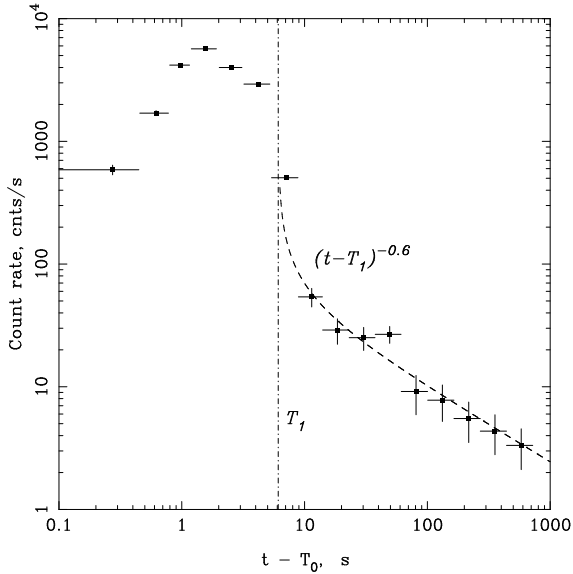}}
\vskip -5mm
\caption{\small The same as  in the fig.3 but for the GRB920723}
\vskip -2mm
\end{figure}

\begin{figure}[th] 
\centerline{\epsfxsize=76mm \epsffile{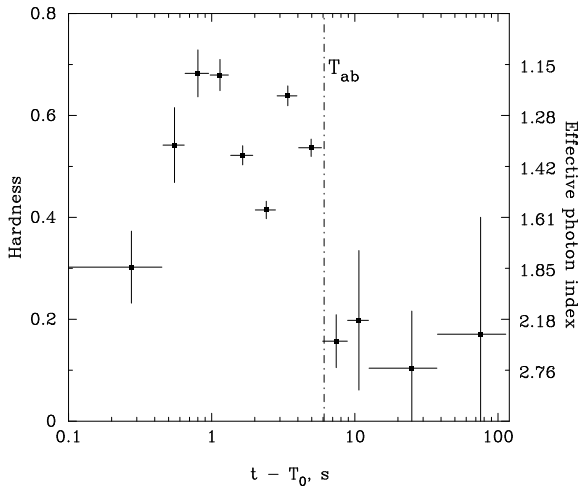}}
\vskip -5mm
\caption{\small The same as in the fig.4 but for the GRB920723 }
\vskip -5mm
\end{figure}

\section{Conclusions and discussion}

The PHEBUS instrument aboard the GRANAT observatory detected soft gamma-ray
afterglow from the GRB~920723 and GRB~910402. These events are the brightest
bursts detected by the PHEBUS instrument. In both cases the light curve of
the
burst makes smooth transition from the main burst into the
afterglow emission (fig.3 and 5).
The best fit power law indices of time for the afterglow are
$-0.70 \pm 0.04$ during $85\--700~{s}$ after GRB~910402 
and $-0.60{\pm}0.05$ during $8-700$~{s} after GRB~920723 (at $1\sigma$ 
confidence level). In both cases the  beginning 
of the afterglow emission ($T_{1}$) coincides within statistical errors 
with the abrupt softening of the emission  ($T_{ab}$).
 Just after GRB event the energy spectrum of the
afterglow  emissions is much
softer than the energy spectrum of the main GRB events.  
The average hardness ratio of the main GRB event (ratio of count rates 
in $250\--800~{keV}$ to $100\--250~{keV}$) is equal to $0.4\--0.5$. This 
corresponds  to the power-law spectrum with the 
photon index of $\sim1.3\--1.5$. After abrupt 
transition to the afterglow both bursts have  much softer
spectra with the photon index of $\sim2.5$.

Thus the moments corresponding to the
beginning of domination of the emission fading as the power law 
during afterglow in both 
cases   coincide
within statistical errors
with the abrupt softening of
the energy spectra emitted by the both  gamma-ray burst sources in soft
gamma-rays $(100\--800~{keV})$.

In the internal/external shock scenario
we can suggest that we found the moment when the much softer 
afterglow emission (connected with external shock) is 
starting to dominate over harder GRB emission
(connected with internal shocks).

During the afterglow of GRB910402 the statistically 
significant hardening of the emission spectra is observed (fig.4).
The afterglow spectrum just after the gamma-ray burst has the photon index of
$\sim2.5$. As one can see from fig.4 after $\sim200~{s}$ the afterglow
photon index was already $\sim1.6$. 
This is the first observation  of hardening of the afterglow 
spectra with time. The possibility of such hardening is under
discussion in different GRB afterglow models.

The afterglow emission intensities in $100\--500~{keV}$ energy range 
are rather faint in comparison with 
gamma-ray bursts.  
In the case of GRB~920723 the  afterglow emission contains 
$\sim 6\%$ of the total burst energy
that was  emitted   during the interval of $6.1\--700~{s}$
in the $100\--500~{keV}$ energy range. 
In the case of GRB~910402 only  $\sim1.6\%$  of the GRB energy 
was released during the observed afterglow in $73.2\--700~{s}$ time interval.

\acknowledgements
This work was supported by the Russian Foundation for Basic Research
(projects no. 96-15-96930 and 00-02-17251).
We thank Marat Gilfanov and Alexey Vikhlinin for helpful comments and 
discussions.
Authors are grateful to referee of this paper, Ed Fenimore, 
for useful remarks.

\end{document}